\documentclass[doublecol]{epl2} 

\title{Nonequilibrium thermodynamics as a gauge theory}

\author{Matteo Polettini \thanks{E-mail: \email{polettini@bo.infn.it}} }

\institute{Dipartimento di Fisica, Universit\`a di Bologna, via Irnerio 46, 40126 Bologna, Italy\\
 INFN, Sezione di Bologna, via Irnerio 46, 40126 Bologna, Italy}
  

\abstract{ We assume that markovian dynamics on a finite graph enjoys a gauge symmetry under local scalings of the probability density, derive the transformation law for the transition rates and interpret the thermodynamic force as a gauge potential. A widely accepted expression for the total entropy production of a system arises as the simplest gauge-invariant completion of the time derivative of Gibbs's entropy. We show that transition rates can be given a simple physical characterization in terms of locally-detailed-balanced heat reservoirs. It follows that Clausius's measure of irreversibility along a cyclic transformation is a geometric phase. In this picture, the gauge symmetry arises as the arbitrariness in the choice of a prior probability. Thermostatics depends on the information that is disposable to an observer; thermodynamics does not.}

\usepackage{amsthm}
\usepackage{MnSymbol}
\usepackage{mathrsfs}
\usepackage{nicefrac}
\usepackage{latexsym}

\newcommand{\be}{\begin{equation}}
\newcommand{\ee}{\end{equation}}
\newcommand{\bea}{\begin{eqnarray}}
\newcommand{\ba}{\begin{array}}
\newcommand{\eea}{\end{eqnarray}}
\newcommand{\ea}{\end{array}}
\newcommand{\de} {\mathrm{d}}
\newcommand{\bs}[1] {#1}

\begin{document}


\maketitle

\section{Introduction}

Open systems subject to dissipation are usually modelled through markovian dynamics and further characterized by nonequilibrium thermodynamics \cite{schnak}. The link between dynamics and thermodynamics is the concept of thermodynamic force. 
In this work we assume that dynamics enjoys a gauge symmetry and show that the thermodynamical sector  of the theory arises very naturally from the requirement that physical observables, in particular the entropy production, are gauge invariant. The driving force plays the role of gauge potential.

A gauge theory has an internal symmetry whose action leaves all physical observables invariant \cite{baez}. Strictly speaking, \textit{any} symmetry leaves observables invariant, there comprising the choice of units and reference frames \cite{tao}. However, conventionally one refers to gauge symmetries as to supplementary redundancies of the variables, which are not related to space-time symmetries: hence ``internal''. 
One copy of the internal space $\Psi_i$, where the gauge symmetry acts, is thought to be attached to each point $i$ of space-time, or more generally of a base manifold. The peculiarity of gauge symmetries is that they act locally, that is, pointwise differently. The tool to compare nearby transformations is the connection, or gauge potential, $A$. 
Its transformation properties are employed to adjust observables (\textit{e.g.} the Lagrangian in field theories) in order to guarantee the internal consistency of the theory. 

{Crucial gauge-invariant observables can be built by circulating the connection along closed paths $\gamma$ over the manifold, obtaining the Wilson loops of the theory
\be
W(\gamma) = \mathrm{tr} \,\left\{ \mathcal{P} \exp \oint_{\gamma} A \right\}.
\ee
The trace over internal group-representation indices and the path-ordering operator $\mathcal{P}$ will be inessential in our theory}.
{Wilson loops} are the finite version of the curvature of a connection, telling how different paths  between two given points carry different transformations. {They} play an ever more prominent role in our understanding of many phenomena besides field theories, from adiabatic phases in quantum mechanics \cite{geometric} and quantum computation schemes \cite{holonomic} {to a proposal of quantum gravity  \cite{lqg}.}

According to a celebrated theory by J. Schnakenberg \cite{schnak}, the macroscopic external forces (affinities) {that} maintain a nonequilibrium system into a dissipative steady state are defined as circuitations of the thermodynamic force. Schnakenberg claimed that affinities are the fundamental observables which characterize nonequilibrium steady states, bringing as evidence that they satisfy Onsager's reciprocity relations in the regime where forces and currents are linearly related. Andrieux and Gaspard \cite{andrieux} complemented his insight with a Green-Kubo relation for the linear response coefficients; the author proved that affinities serve as constraints for a formulation of the minimum entropy production principle near equilibrium \cite{polettini}. 

Schnakenberg's theory applies to markovian evolution of a normalized probability density over a discrete {phase  space}, obeying the master equation
\be \dot{\rho}_i(\tau) =  \sum_j J_{ij}(\tau) =  \sum_{j} \big[  w_{ij} \rho_j(\tau) -  w_{ji} \rho_i(\tau) \big] \label{eq:master}\ee
where $w_{ij}$ is a positive transition rate from vertex $j$ to vertex $i$, and $J_{ij}= w_{ij} \rho_j -  w_{ji} \rho_i$ is the probability current. In spite of space-time, the base manifold is the discrete phase space $V$ consisting of a number $|V|$ of vertices. We define for later convenience an order relation $i < j$. Microstates are pairwise connected by a number $|E|$ of edges $ij \in E$, which are assigned a conventional orientation $ij = i \gets j$. We assume that the graph $G = (V,E)$ is connected and that forward and backward rates are nonnull along all edges. The prominent role played by circuitations in Schnakenberg's theory is an indicator that it is a gauge theory, with the driving force
\be
A_{ij} =  \log \frac{w_{ij}}{w_{ji}} \label{eq:driving}.
\ee
playing the role of the gauge potential. To our knowledge, until very recently the geometric nature of the thermodynamic force was confined to the mathematical literature \cite{qiansbook} and to work by Graham \cite{graham}, with no explicit reference to gauge invariance. In the year of writing, for continuous  diffusive processes the interpretation of the force as a gauge potential has been put forward by Feng and Wang \cite{feng}. Sagawa and Hayakawa \cite{sagawa} made a proposal for a gauge potential connecting nonequilibrium steady states along slowly driven protocols; differently from Feng and Wang's, their connection has null curvature. They also observe that ``the gauge symmetry does not seem to play any important role''. We fill the gap, taking an orthogonal approach: we do not assume the connection to be given, and derive it as the most natural candidate which guarantees covariance of the master equation; the appearence of a gauge potential is a byproduct of the symmetry, a conclusion which in a way parallels Abe and Kaneko's analysis of driven quantum equilibrium states \cite{abe}.

\section{Priors} Prior to the derivation, we provide a physical and slightly phylosophical motivation for the gauge symmetry, with a detour into continuous variables. In the ``informationist'' approach to statistical mechanics, whose forefather is Jaynes \cite{jaynes}, the Gibbs-Shannon entropy
\be
S(\rho) = - \sum_i \rho_i  \log \rho_i
\ee
is a measure of the ignorance that an observer has about the state of the system. Maximization of the entropy, subject to constraints according to whatever pieces of information the observer gains from measurement, produces the most plausible distribution \textit{given that} the vertices of the system are a priori equally likely, according to Laplace's principle of insufficient reason. There is a source of subjectivity, which Jaynes accepted as physical, related to the choice of the observables one sets up to measure. However, there is a second one which is in-built and  which made Jaynes uneasy in his earlier writings. From \cite{jaynes}:
\begin{quote} \small
Laplace's ``Principle of Insufficient Reason'' was an attempt to supply a criterion of choice [\ldots] However, except in cases where there is an evident element of symmetry that clearly renders the events ``equally possible'', this assumption may appear just as arbitrary as any other that might be made. Furthermore, it has been very fertile in generating paradoxes in the case of continuously variable random quantities, since intuitive notions of ``equally possible'' are altered by a change of variables.
\end{quote}
He then advised to replace Laplace's with the maximum entropy principle. So doing, he swept the dirt under the carpet, as the {Shannon-Khintchin's set of axioms for the entropy include equiprobability  \cite{suyari}, and Shannon's monotonicity axiom \cite[p. 630]{jaynes} makes reference to it}.

Moreover, alleged paradoxes are found in the continuous variables case, where the differential entropy \cite[Ch.9]{cover}
\be
S(\rho) = - \int_X \de \bs{x}\; \rho(\bs{x})  \log \rho(\bs{x}) \label{eq:gibbsent}
\ee
has been a source of dismay \cite{hnizdo,vulpiani,white}, for it is not invariant under a change of variables. In fact, letting $\bs{x}' \mapsto  \bs{x}(\bs{x}')$ be invertible with jacobian $J(\bs{x}') = \det (\partial \bs{x} / \partial \bs{x}')$, one has
\be
S(\rho) = S(\rho') + \langle {\, \log\,} J \, \rangle_{\rho'} \label{eq:entropycon}
\ee 
where the probability measure and its density  transform respectively as a volume form and as a volume density,
\be
 \rho'(\bs{x}')\de \bs{x}' = \rho(\bs{x})  \de \bs{x} ,\qquad \rho'(\bs{x}') = J(\bs{x}') \rho \left(\bs{x}(\bs{x}')\right).  \label{eq:tdc}
\ee
Related to this is the following riddle (from \cite[Ch.8]{riddles}): if we pick a number $x$ between $1$ and $10$ \textit{at random}, the probability that it is smaller than $5$ is $\nicefrac{1}{2}$; but if we pick $x'$ at random between $1$ and $100$, the probability that it is smaller than $25$ is $\nicefrac{1}{4}$. How is it possible that picking either a number or its square aren't equally likely? The solution to this puzzle is to recognize that the choice of an arbitrary prior is congenital. It hides in that ``at random'' which is the continuous counterpart of Laplace's principle: in the first case we assume $x$ is uniform, so that the prior is $\nicefrac{1}{10} \,\de x$; in the second we assume that $x'=x^2$ is uniform, with prior $\nicefrac{1}{100}\,\de x' = \nicefrac{1}{50} \,x\, \de x$. Formally, in order to make eq.(\ref{eq:entropycon}) mathematically sound, one will interpret $\rho(\bs{x})$ as the Radon-Nicodym derivative of the probability measure $\rho(\bs{x})\de \bs{x}$ with respect to the arbitrary prior $\de \bs{x}$. A change of variables corresponds to a change of prior. 

This is not, as Jaynes thought, an artifact of continuous variables. Think for example of a dice. Basing on visual impressions --- which, by the way, are the result of a measure process --- we might be tempted to assign equal probabilities $\nicefrac{1}{6}$ to each face. However, if we knew that  an incredibly huge mass was hidden near one of its corners, due to friction with air and the inelastic impact with the gaming table, we would have sufficient reason to believe that the three faces which are adjacent to the {loaded} corner will have probability approximately $\nicefrac{1}{3}$, and the others near zero. Our gambling strategy will depend on this prior knowledge. As a way out, in spite of invoking measure theory, up to additive constants
 we might just regard (neg)entropy as a special case of relative entropy
 \be
S(\rho\, ||\, \mu) = \sum_{i} \rho_i  \log \rho_i / \rho^{\mathrm{(pr)}}_i,
\ee
with respect to a uniform prior $\rho^{\mathrm{(pr)}}_i = |V|^{-1}$.
We refer the reader to Banavar and Maritan's work \cite{maritan} for some nice physical implications of working with relative entropy. 

The physical rationale is that the quantification of the entropy of a system depends on the choice of the underlying degrees of freedom. If we assume that all configurations of positions and momenta of a number of classical particles are equiprobable, we implicitly coarse-grain the atomic and subatomic structure. The question ``how much entropy is within a body'' makes no sense on its own, since we can always go deeper into the inner structure of matter, according to the resolution of our ``gedanken-apparatus''. \textit{Thermostatics} depends on the prior.
However --- and here comes the key point --- if we put a gas in contact with heat reservoirs, the process will occur in exactly the same manner, irregardless of our quantification of the system's entropy. \textit{Thermodynamics} is independent of the prior.

This is the gauge principle we assume. A gauge transformation is a change of priors. Gauge fixing means to choose a prior; it is analogous to the choice of a position with respect to which we measure displacements, with the important difference that in Newtonian mechanics the choice of a reference frame is absolute, in gauge theories the choice of a ``reference frame'' varies point by point. 

\section{Gauge transformations}
With the r.h.s. of eq.(\ref{eq:tdc}) in mind, we postulate that
\be \rho'_i  \,=\, e^{-\varphi_i} \rho_i \label{eq:transdens} \ee
is a symmetry of the theory, with $\varphi_i \in \mathbf{R}$. For sake of consistency, we need to prescribe transformation laws for all of the objects which partecipate to the master equation.

First, consider transition rates. We assume an edgewise and linear transformation law
$w'_{ij} \,=\, v_{ij} w_{ij} e^{\varphi_j}$,
where we singled out $e^{\varphi_j}$, without loss of generality.  {Notice that the special case with all $v_{ij}=1$, namely
\be
w'_{ij} = w_{ij} e^{\varphi_j} \label{eq:transtrans}
\ee
leaves the currents invariant,
\be
J'_{ij} \; = \; J_{ij} \label{eq:transcur},
\ee
The following graph-theoretical analysis proves that eq.(\ref{eq:transtrans}) is the most general edgewise linear transformation law compatible with eq.(\ref{eq:transdens}).}

Under the assumptions of connectedness and of nonvanishing rates stated above, there exists a unique steady state  $\rho^\ast$ of the master equation, which makes the r.h.s. of eq.(\ref{eq:master}) vanish.
An explicit expression for $\rho^\ast$, due to Kirkhhoff, is known \cite{schnak,zia}: we shall use it to constrain {the $v_{ij}$'s}. 
The {recipe} goes as follows. (i) Consider a rooted spanning tree $T_i$, \textit{i.e.} a maximal set of edges with no cycles and with  one preferred vertex $i$ chosen as the root; all edges are oriented so that there exists exactly one directed path from any other vertex to $i$. (ii) Take the product of the transition rates along its edges,
\be\pi_w(T_i) \,=\, \prod_{jk\, \in \,T_i} w_{jk} .\nonumber \ee
(iii) Form the sum over all possible rooted spanning trees to obtain the polynomial $Z_i = \sum_{T_i} \pi_w(T_i)$, which is homogeneous of degree $|V|-1$ in the transition rates. (iv) The steady state is
$\rho^\ast_i \,=\, Z_i / \sum_j Z_j$. By construction, each vertex of the graph, but $i$, is the starting point of exactly one edge of $T_i$, so that we can factorize
\be
Z'_i \,=\,  e^{\phi-\varphi_i}   \sum_{T_i} \pi_v (T_i) \pi_{w}(T_i),
\ee
where $\phi = \sum_{j\in V} \varphi_j$. By eq.(\ref{eq:transdens}), $Z'_i$ must be proportional to the analogous expression for $e^{-\varphi_i} Z_i$. We obtain
\be
\sum_{T_i} \pi_{w}(T_i) \,=\,  c^{1-|V|} \sum_{T_i} \pi_v (T_i) \pi_{w}(T_i) 
\ee
where $c$ is a proportionality constant. Since all transition rates are positive and the equality must hold $\forall w$'s , if follows that $\pi_v (T_i) = c^{|V|-1}$ independently of the spanning tree.
Furthermore, the transformation law should be universal, \textit{i.e.} it should not depend on the specific graph. As graphs become larger, tipically the number of spanning trees grows exponentially in the number of edges of the graph ---whereby ``tipically'' loosely means ``for most graphs'' \cite{janson}. This entails that the number of equations specified by $\pi_v (T_i) = c^{|V|-1}$ becomes enormously larger than the number of the unknowns. The only universal solution is $v_{ij} = c , \forall \, ij$; the constant can then be scaled to unity with a redefinition of the time unit $\tau \to \tau/c$. {$\Box$}

We now face a seeming paradox. In fact, considering the transformed master equation $\dot{\rho}'_i  = \sum_j  J'_{ij}$
and keeping into account eq.(\ref{eq:transdens}) and  eq.(\ref{eq:transcur}), we obtain an equation which is \textit{not} equivalent to the starting master equation, eq.(\ref{eq:master})! The solution delves into the geometrical interpretation of summation symbols. We introduce the incidence matrix
\be
^1\hspace{-0.05cm}\partial_{\,i}^{\,jk} =  \left\{ \begin{array}{ll} + 1, &~~ \mathrm{if}~
 j < k, \;k = i  \\ 
-1,  &~~ \mathrm{if}~
  j < k, \;j = i \\
0, & ~~\mathrm{elsewhere}
\end{array} \right. ,
\ee
and rewrite the master equation as a continuity equation
\be
\dot{\rho} \,+\;  ^1\hspace{-0.05cm}\partial  J \,=\, 0\label{eq:continuity}.
\ee
Technically speaking, the incidence matrix is a boundary operator which maps edges into their boundary vertices. Normalization of the probability can be written as
\be
 \sum_{i \in V} \rho _i  = \, ^{0}\hspace{-0.05cm}\partial  \rho  =  1 \label{eq:normalization}
\ee
where we introduced one further boundary operator $ ^{0}\hspace{-0.05cm}\partial = (1,1,\ldots,1)$, which maps vertices to the connected component of the graph they belong to. Although this latter definition is rather trite, it gets more interesting when the graph has several disconnected components. 
Notice that $^{0}\hspace{-0.05cm}\partial \,^{1}\hspace{-0.05cm}\partial = 0$, from which conservation of probability follows. Therefore ``summation over $i$'' has different geometrical meanings according to the context.

Strictly speaking, $\rho_i$ should not be considered as a number, but rather as a one-component vector which lives in the internal vector space $\Psi_i \cong \mathbf{R}$ which is attached to vertex $i$. The gauge transformation eq.(\ref{eq:transdens}) is interpreted as a linear change of basis in $\Psi_i$. It follows that we should consider the boundary operator's entries as linear maps on $\Psi_i$, which therefore transform according to  
\be
{^{1}\hspace{-0.05cm}\partial}'_{\,i}  \;=\; e^{-\varphi_i}\; ^{1}\hspace{-0.05cm}\partial_i , \qquad
{^{0}\hspace{-0.05cm}\partial}'_{\,i} \;=\; e^{\varphi_i} \; ^{0}\hspace{-0.05cm}\partial_i .
\ee
With this prescription, eq.s (\ref{eq:continuity}) and (\ref{eq:normalization}) are covariant. To simplify the notation a bit, we introduce a modified sum symbol $\sum'$ such that
{\bea
{\sum_i}'  = \sum_i e^{\varphi_i}
\eea}
This modified symbol is crucial for the up-coming result, so let us further linger on it. Consider the average of a gauge-invariant vertex function $f' =f$ (a scalar field),
\be \langle f \rangle_\rho = \sum_i \rho_i f_i = {\sum_i}' \rho'_i f_i . \ee
Requiring gauge invariance $\forall f$ yields the transformation law for the summation symbol. In other words, while the probability measure  $\langle\; \cdot \;\rangle_\rho$ is gauge invariant, the probability density  $\rho_i$ is not, in analogy with the continuous variables case, see eq.(\ref{eq:tdc}).

We now focus on the Gibbs-Shannon entropy, which transforms according to
\be
\delta S \,=\, S'[\rho'(\tau)] - S[\rho(\tau)] \,=\, \langle \varphi \rangle_\rho  \,=\, - S(\rho'\,||\,\rho)
\ee
where $S'$ is calculated using $\sum'$. On the r.h.s., the transformation law is succinctly expressed 
in terms of relative entropy. Remarkably, while relative entropy is not a difference of entropies, in this context it is naturally interpreted as (minus) the entropy change after a gauge transformation.
The rate at which the entropy of the system changes is subject to
\be \delta \dot{S} \, = \,\sum_{i < j} J_{ij} (\varphi_i - \varphi_j). \label{eq:transent} \ee
In gauge theories, non-gauge invariant terms are adjusted with the introduction of a connection, which is an antisymmetric edge variable $A_{ij}=-A_{ji}$ such that
\be \delta A_{ij} =  \varphi_j - \varphi_i. \label{eq:transconn}  \ee
Once a connection is given, the term\footnote{In this case, the sum has meaning of a bilinear form from the space of edges to real numbers, so no gauge transformation is needed.}
\be
\sigma = \sum_{i < j} J_{ij} A_{ij}
\ee
has a transformation law which balances eq.(\ref{eq:transent}), making $\dot{S} + \sigma$ invariant. In principle, connections can be constructed as convex linear combinations of terms such as
\be
 \log \frac{\rho^\ast_{i}}{\rho^\ast_{j}}, \quad  \log \frac{\omega_{j}}{\omega_{i}}, \label{eq:choices}
\ee
where $\omega_i = \sum_{k} w_{ki}$ is the average frequency of a jump out of site $i$. So, for example, adding $\sum_{i < j} J_{ij}  \log  \rho^\ast_{i} / \rho^\ast_{j}$
yields the relative entropy with respect to the steady state. The latter plays an important role in the theory of Markov processes as a Lyapunov functional \cite[Sec. V]{schnak}; fitly, it is gauge invariant, while entropy \textit{per se} is not.  However, the options listed above are, technically speaking, exact: they are differences of vertex functions, so that their circuitations vanish, thus making the graph's geometry rather dull. As a further consequence, gauge invariant terms obtained this way vanish at the steady state.

A good candidate as a ``truly edge'' connection variable is given by the driving force, defined in eq.(\ref{eq:driving}). Although it is not the only antisymmetric edge variable that one could engineer which transforms according to eq.(\ref{eq:transconn}), it is certainly the simplest. Then $\dot{S} + \sigma$ coincides with 
Schnakenberg's total entropy production \cite[eq.(7.6)]{schnak},
\be
\sigma_{tot} \;=\; \dot{S} + \sigma \;=\; \sum_{i < j} J_{ij}   \log \frac{w_{ij}\rho_j}{w_{ji} \rho_i},
\ee
which is widely accepted as \textit{the} entropy production rate of a Markov process \cite{seifert,maes}. 
In this setting $\sigma$ arises as the simplest term which completes $\dot{S}$ into a gauge invariant quantity and which does not vanish at the steady state.

A gauge transformation will result in a shift of a total time derivative from $\sigma$ to $\dot{S}$, with a consequent redefinition of the internal entropy and of the entropy flow towards the environment. For example, letting $\varphi_i =  \log \rho_i^\ast$, we obtain 
\be
{S' = - S(\rho\, ||\, \rho^\ast)} , \quad \sigma' = \sum_{i < j} J_{ij}  \log \frac{w_{ij} \rho_j^\ast}{w_{ij} \rho_j^\ast} ,
\ee
whose microscopic analogues along single stochastic trajectories have been interpreted by Esposito and Van den Broeck  as non-adiabatic and adiabatic terms, obeying detailed fluctuation theorems  \cite{esposito2}. Gauge transformations of fluctuation theorems will be discussed in a later work.

\section{Wilson loops}

From a geometrical viewpoint \cite{ddg}, not only $A$ provides a connection over the manifold, but  it also constitutes a measure of the oriented lenght of paths along chains of edges $\eta = (i_n i_{n-1},\ldots,i_1i_0)$,
\be
\Sigma(\eta) \,=\, \sum_{\kappa=1}^{n} A_{i_{\kappa}i_{\kappa-1}} \,=\, \int_\eta A \, \label{eq:entprod}.
\ee
Since the lenght is additive upon composition of paths, the real positive numbers obtained by exponentiating $\Sigma(\eta)$ can be thought of as elements in the multiplicative group of real positive numbers $(\mathbf{R}^+,\times)$, which is the gauge group of the theory. In the representation theory of groups, group elements are not seen as ``static'' objects, but rather as  ``active'' linear maps; they act on vectors $\psi _i$ which live in the internal vector spaces $\Psi_i$. Such vectors acquire phases as they are parallel transported along paths, thus connecting far-apart vertices,
\be
\underline{\psi}{}_{i_n} = \exp {\Sigma(\eta)} \; \psi_{i_0}, \label{eq:partrans}
\ee
where $\underline{\psi}{}_{i_n}$ represents the result of parallel transport along path $\eta$. In our case, due to the very simple gauge group, the displaced vector is just a real number and parallel transport produces a scaling factor.  The intepretation of group elements as linear maps further entails that new equivalent representations can be obtained by performing basis transformations in $\Psi_i$, one per each vertex: this yields a gauge transformation. In the case at hand, such a basis change amounts to {an orientation-preserving} rescaling $\psi'_i = e^{-\varphi_i} \psi_i$. Transformed vectors are parallel transported according to $\underline{\psi}'{}_{i_n} = \exp {\Sigma'_\eta} \; \psi'_{i_0}$, where $\Sigma'$ is a new representation of the group element, defined in terms of a transformed connection $A'$. Requiring equivalence with eq.(\ref{eq:partrans}) {for any possible path $\eta$} yields the transformation law for the vector potential{, eq.(\ref{eq:transconn}).}
{Grossly, this introduces the geometrical framework for gauge theories.} 

Gauge transformations define an equivalence relation ``\,$\sim$\,'' between gauge potentials; so, for example, the adiabatic force $ \log (w_{ij} \rho_j^\ast / w_{ji} \rho_i^\ast)$ \cite{esposito2} is gauge-equivalent to ours, eq.(\ref{eq:driving}). The connection is said to be exact when it is equivalent to $A'_{ij} = 0$.
It is well known \cite{graham,qiansbook,feng,qian} that equilibrium systems are characterized by an exact potential. In fact, when $A_{ij} = \varphi_i - \varphi_j$, the steady solution of the master equation is $\rho^\ast_i \propto e^{-\varphi_i}$, as  direct substitution into eq.(\ref{eq:master}) shows. Detailed balance follows: 
\be \frac{w_{ij}}{w_{ji}} = \frac{\rho^\ast_i}{\rho^\ast_j}.\ee

Along closed cycles $\gamma$, with $i_0 = i_n$, the exponentiated lenght is a Wilson loop.
When Wilson loops are all unity the connection is exact and the oriented lenght of an open path only depends on the extremal vertices, and not on the particular path which connects them, for which reason the connection is said to be flat. As a remarkable consequence, Kolmogorov's criterion \cite{kolmogorov,zia}
is equivalent to all Wilson loops being equal to unity. Hence detailed balanced systems can be seen as the special class of models with a flat connection, with zero curvature; they all belong to the same equivalence class.


On a discrete state space, knowledge of a finite number of Wilson loops suffices to characterize the connection. The so-called Mandelstam identity
\be W({\gamma_1 \circ \gamma_2}) = W({\gamma_1}) W({\gamma_2})\ee
 allows to compose loops.  A basis of loops can be found this way. Consider an arbitrary spanning tree $T$ of the graph --- this time with no preferred root and orientation. Let $i_{\alpha} j_{\alpha}$ be one of the edges which do not belong to $T$. By definition, adding $i_{\alpha} j_{\alpha}$ to the spanning tree generates a cycle $\gamma_\alpha$, which can be oriented according to the orientation of $i_{\alpha} j_{\alpha}$. By Euler's formula, there are $|E|-|V|+1$ such cycles. It is a basic graph-theoretical  result that any loop can be decomposed in terms of the $\gamma_{\alpha}$'s \cite[Part I, Ch.4]{biggs}. 
Let  {$e_{ij}^\alpha$  be $+1$ if $ij=i_{\alpha} j_{\alpha}$, $-1$ if $ji = i_{\alpha} j_{\alpha}$, otherwise it is zero.} It can be shown that
\be
 \log \prod_{\alpha}  W({\gamma_{\alpha}})^{{e_{ij}^\alpha}} ~\sim~ A_{ij} . \label{eq:equiv}
\ee
Hence Wilson loops allow to reconstruct the gauge potential, up to gauge transformations \cite{giles}. By eq.(\ref{eq:equiv}), the choice of a spanning tree fixes the gauge by selecting one particular representative in the equivalence class of $A_{ij}$.

Spanning trees also allow to give a physical interpretation of the connection, as follows. Any graph which coincides with a spanning tree, $E=T$, has no cycles, hence it can only accomodate equilibrium systems. Then there exists a vertex function $\varphi_i = \beta u_i$ such that
\be
w_{ij} / w_{ji} \,=\, e^{\beta(u_j - u_i)}, \qquad ij \in T,
\ee
where we introduced an inverse temperature $\beta$, in units of Boltzmann's constant. The inverse temperature and the energy $u_i$ are determined up to an energy shift and a rescaling of units, $u_i \to k( u_i + v )$, $\beta \to k^{-1} \beta$.  In general, adding further edges $i_{\alpha} j_{\alpha}$ to the graph will not result in a detailed balanced system, unless we fine-tune their rates. We then define a new set of temperatures $\beta^\alpha$, such that
\be
w_{i_\alpha j_\alpha} / w_{j_\alpha i_\alpha} \, = \, e^{\beta_\alpha(u_{j_{\alpha}} - u_{i_{\alpha}})}, \qquad i_{\alpha} j_{\alpha}	\in E\setminus T.
\ee
We just proved that the thermodynamics of any collection of transition rates can be described in terms of at most $|E|-|V|$ reservoirs, each at its own temperature, satisfying the condition of local detailed balance \cite{esposito}. In this ``minimal'' case each transition is due to the interaction with exactly one reservoir. This \textit{ansatz} allows to recast the basis Wilson loops in this form
\be
W(\gamma_\alpha) \,=\,  \exp \left[ (\beta - \beta_\alpha) ( u_{i_\alpha} - u_{j_\alpha} ) \right] .
\ee
Therefore, temperature differences are the fundamental thermodynamic forces of nonequilibrium systems, as one could expect. Since there is no external time-dependent driving, which would result in time-dependent transition rates, no work is performed by an external agent along one single realization of the process, and by the first law of stochastic thermodynamics \cite{seifert}, along a transition the energy gap $\delta u$ coincides with the heat exchanged $ \delta q$. It is then illuminating to rewrite the geometric phase as
\be
 \log W(\gamma) \,=\, \oint_\gamma \frac{\delta q}{T},
\ee
yielding Clausius's measure of irreversibility along one realization of a cyclic irreversible process. The lenght $\Sigma(\eta)$ is the entropy exchanged with the environment along any trajectory which performs a sequence of jumps,
whichever the jumping times might be. This notion is completely independent of the time parametrization of the trajectory: it is purely geometrical.

\section{Final remarks}

The above construction can be easily generalized to Markov processes with time-dependent transition rates and to time-dependent gauge transformations. In this respect, our formalism has evident points of contact with stochastic pumping along cyclic protocols, whose geometrical nature has been recently studied \cite{pump}.  It would be a conceptual advance to give a unified description of both aspects of NESM. We notice in passing that Sinitsyn \cite[\S 6]{sinitsyn} makes a remark on gauge transformations applied to the current generating function, {arguing that they follow from the modification of the  ``prior''  currents which have flown before a given initial time.}

Regarding the nature of gauge transformations, for continuous variables they have been shown to follow from coordinate changes. Thus the gauge group could be seen as (a subgroup of) the group of diffeomorphisms. It has been a matter of disagreement \cite{weinstein}\cite[\S 2.1.3]{lqg} whether diffeomorphisms and gauge transformations should be considered by the same standards; the diatribe mainly revolves around gravity and its formulation as a local affine theory \cite{hehl}. The identification of ours as a gauge transformations is justified by the usage of the gauge machinery, which is analogous to well-{established} practice {for the formalization of geometric phases in QM and of electromagnetism as a $U(1)$ gauge theory. Employing analogies with the latter, C. Timm \cite{timm} discussed a slightly different gauge-theoretic structure for master equations.}

To conclude, while we are conscious that the very simple gauge group makes the geometrization of irreversible thermodynamics unnecessary for all practical purposes, it allows to better appreciate the importance of macroscopic affinities as fundamental observables, and it might serve as a good starting ground for later generalizations. We point out that a Schnakenberg-type analysis is still lacking for quantum nonequilibrium systems, either described by a Lindblad-type equation or a by a more general interaction of a system with reservoirs of quantum degrees of freedom. It is tenable that excursions to the quantum world might require more interesting gauge groups and a more pertinent application of gauge theory.

\acknowledgments

Joint work with Massimiliano Esposito is at the basis of the physical interpretation of loops; the author is also grateful for his useful comments. {Stimulating discussion with A. Vulpiani and A. Montakhab is recognized.}

\end{document}